\documentclass[twocolumn,showpacs,amsmath,amssymb,superscriptaddress]{revtex4}
\usepackage{amsmath,amssymb,color,latexsym,natbib,graphicx}
\usepackage{dcolumn}
\usepackage{bm}
\def\Final#1{{\textcolor{black}{#1}}} 
\newcommand{\p} {\partial}
\def\eg{{\it e.g.}\ } 
\def\ie{{\it i.e.}\ }

\def\ww{{\bf w}}
\def\vv{{\bf v}}

\def\d_M{{\bf d_M}}

\def\rr{{\bf r}}

\def\jj{{\bf j}}
\def\bb{{\bf b}}
\def\aa{{\bf a}}
\def\xx{{\bf x}}
\def\UU{{\bf U}}

\def\bOmega{{\boldsymbol \Omega}}

\def\be{\begin{equation}}
\def\ee{\end{equation}}
\def\ba{\begin{eqnarray}}
\def\ea{\end{eqnarray}}

\def \pmbtext#1{\leavevmode
     \setbox0\hbox{#1}
     \kern0,4pt \copy0 \kern-\wd0
     \kern-0,2pt \raise0,3pt \box0 }
\begin{document}

\preprint{1}

\title{\Final{Chiral Exact Relations for Helicities in Hall Magnetohydrodynamic Turbulence}}
\author{Supratik Banerjee}
\email{supratik.banerjee@uni-koeln.de}
\affiliation{Universit\"at zu K\"oln, Institut fur Geophysik und Meteorologie, Pohligstrasse 3, 50969 K\"oln, Germany}
\author{S\'ebastien Galtier}
\email{sebastien.galtier@lpp.polytechnique.fr}
\affiliation{LPP, \'Ecole polytechnique, F-91128 Palaiseau Cedex, France}
\affiliation{Univ. Paris-Sud, Universit\'e Paris-Saclay, France}

\date{\today}

\begin{abstract}
Besides total energy, three-dimensional incompressible Hall magnetohydrodynamics (MHD) possesses two inviscid invariants which are 
the magnetic helicity and the generalized helicity. New exact relations are derived for homogeneous \Final{(non-isotropic)} stationary Hall MHD 
turbulence (and also for its inertialess electron MHD limit) with non-zero helicities and in the asymptotic limit of large Reynolds numbers. 
The universal laws are written only in terms of \Final{mixed} second-order structure functions, \Final{\ie the scalar product of two different 
increments. It} provides, therefore, 
a direct measurement of the dissipation rates for the corresponding invariant flux. 
This study shows that the generalized helicity cascade is strongly linked to the left polarized fluctuations while the magnetic helicity cascade is 
linked to the right polarized fluctuations. 
\end{abstract}
\maketitle

\section{Introduction} \label{intro}
Hall MHD is a mono-fluid plasma model appropriate for probing \Final{some of the} physical processes 
\Final{(other than pure kinetic effects)} at length scales smaller than the scales of standard MHD. 
Unlike MHD, this model introduces a decoupling between the electrons and the ions {\it via} the so-called Hall term in the generalized Ohm's law.
The Hall effect comes to be relevant when the typical length scales are smaller than the ion inertial length $d_i$ ($d_i \equiv c / \omega_{pi}$
with $c$ the speed of light and $\omega_{pi}$ the ion plasma frequency). In a plasma for which the Alfv\'en mode controls the dynamics (incompressible or a 
collisionless plasma \citep{Belcher71}), one can associate this length scale to the ion cyclotron angular frequency $\omega_{ci} = V_A / d_i$ 
(with $V_A$ the Alfv\'en speed) and therefore the Hall MHD regime is valid for the time scales of the order of, or shorter than the ion cyclotron period ${2 \pi }/\omega_{ci}$. 

In space physics, the potential importance of Hall MHD is recognized for a range of phenomena varying from collisionless reconnection, disruption 
of Alfv\'enic filaments, to small-scale solar wind turbulence \cite{Ghosh,Bhattacharjee2004,Dreher,mininni2005,jltp,Mininni2007,Sahraoui07,alexandrova09,Kiyani}.
Despite this fact, fewer number of analytical studies have been performed in Hall MHD \citep{yoshida,galtier08,Araki2015a,Araki2015}, while the standard MHD remains the 
subject of active theoretical and numerical research for more than three decades specially in the framework of turbulence 
(see \eg \cite{pouquet76,galtier00,muller,Buchlin2003,Alexakis,Matthaeus2009,Galtier2012,beres,Banerjee,Meyrand2015}). 
A few direct numerical simulations, however, have been performed in Hall MHD turbulence to study either in a classical way the behavior of the velocity
and magnetic field fluctuations \citep{Ghosh,Mininni2007,Shaikh2009,Miura2014,Stawarz2015}, or in a more sophisticated way by separating the fluctuations in terms 
of polarities (left and right) \citep{Meyrand12}. Shell models have also been used to investigate in particular the transition from standard to Hall
MHD \citep{galtier07,hori,pandit}. 

The Hall MHD model gets much simplified in the limit of small length scales ($k d_i \gg 1$) where the ions are considered as a motionless neutralizing
background and the electron flow practically determines the electric current. This particular limit of Hall MHD, called (the inertialess) electron MHD, 
is much easier to analyze specially in the case of turbulence \citep{kingsep,goldreich92,meyrand10,diamond11}. 
\Final{However, despite being a simple mathematical limit of Hall MHD, electron MHD is physically different from the former as the corresponding plasma dynamics 
is not governed by the force balance 
\cite{Lyutikov13}.} Besides having important roles in laser 
plasmas \citep{sentoku} or in magnetic field reconnection \citep{bulanov,drake}, electron MHD is also relevant for the understanding of the magnetic 
fluctuations in the collisionless solar wind at sub-ion scales (see \eg \citep{Meyrand13}). 
Direct numerical simulations of isotropic electron MHD turbulence show that the magnetic energy spectrum scales like $k^{-7/3}$ \citep{biskamp96,ng03}.
A plausible explanation for this spectrum has been provided by a heuristic model \`a la Kolmogorov \citep{biskamp96} which turns out to be dimensionally
compatible with an exact relation derived for third-order correlation functions \citep{gelectron}. However, the final form of this exact relation has not
been reduced exclusively in terms of two-point fluctuations which weakens any spectral prediction for electron MHD turbulence. 

In addition to the total energy, Hall MHD permits two other inviscid invariants: the magnetic helicity $H_M = \aa \cdot \bb $, where $\bb$ is the magnetic field 
(which will be normalized to a velocity hereafter) and $\aa$ its (normalized) vector potential, and the so-called generalized helicity 
$H_G = \left( \aa + d_i \vv \right) \cdot \left( \bb + d_i \ww \right) $, where $\vv$ is the velocity and $\ww = \nabla \times \vv$ the vorticity 
vector \cite{woltjer58,turner86,Yoshida02,Krishnan04,galtier06}. In the electron MHD limit, we have $\vv = 0$ which denotes $ H_M \equiv H_G$.
Hence, electron MHD allows two invariants: the magnetic energy (which is equivalent to its total energy) and the magnetic helicity. The role of the magnetic
helicity in electron MHD turbulence was investigated recently with three-dimensional direct numerical simulations (with a mean magnetic field) which 
revealed that the propagation of one wave packet moving in one direction leads to the energy transfer towards larger scales \citep{cho11,cho15}. 
This effect interpreted as an inverse cascade shows that one dispersive wave packet may produce another wave packet moving in the opposite direction 
as a result of the magnetic helicity conservation which in turn leads to an inverse cascade. 
The impact of the magnetic helicity was investigated analytically in incompressible MHD \cite{PGP2003}. 
An isotropic Kolmogorov-like exact relation was derived but the final exact relation cannot be expressed only in terms of two-point fluctuations 
thereby rendering the possibility of turbulent spectral prediction less evident. 
The role of the magnetic and generalized helicities in Hall MHD turbulence has never been studied analytically in great detail although it is believed 
to be important \eg for the study of the plasma dynamo \cite{mininni2005} for which a novel experiment has been designed recently \cite{Forest14}. 

In this paper, we derive a new class of exact relations corresponding to the magnetic helicity and generalized helicity conservation in homogeneous 
Hall MHD turbulence (isotropy is not assumed). 
In section II we present the Hall MHD equations; in section III we derive the exact law for the magnetic helicity cascade while in section IV the 
derivation is made for the generalized helicity cascade. 
In section V we \Final{discuss the chiral nature of the helicity cascades} 
and we conclude the paper with a discussion in the last section. 

\section{Incompressible Hall MHD equations} \label{sec2}

The magnetic helicity represents a quantitative measure of the self-linkage of the magnetic field, likewise the generalized helicity represents a quantitative 
measure of the self-linkage of the generalized vorticity ${\bf \Omega} = \bb + d_i \ww$. Below, we shall derive exact relations for {\Final{average}} two-point fluctuations 
\Final{(\ie increments)} 
corresponding to the conservation of the magnetic helicity and generalized helicity in incompressible Hall MHD for which the basic ideal and inviscid equations are
\begin{align}
\p_t{\vv} &= - \vv \cdot \nabla \vv - \nabla P + \jj \times \bb + {\bf f_v} , \label{first1}\\
\p_t{\bb} &= \nabla \times \left( \vv \times \bb\right) - d_i \nabla \times \left( \jj \times \bb\right) + {\bf f_b} , \label{first2}\\
\nabla \cdot \vv &= 0 ,\\
\nabla \cdot \bb &= 0 ,
\end{align}
where $P$ is the fluid pressure, $\jj = \left( \nabla \times \bb\right)$ denotes the current density, ${\bf f_v}$ and ${\bf f_b}$ are stationary
forcing terms (further properties will be given below). Equations (\ref{first1})--(\ref{first2}) can easily be transformed into
\ba
\p_t \ww &=& \nabla \times \left[\vv \times \ww + \jj \times \bb \right] + {\bf f_w} , \label{second1} \\
\p_t \aa &=& \left( \vv - d_i \jj \right) \times \bb + 2 \nabla \psi +  {\bf f_a} ,  \label{second2}
\ea
where ${\bf f_w}$ and ${\bf f_a}$ are the forcing terms and $2 \nabla \psi$ corresponds to an arbitrary choice of gauge.

In the previous equations the dissipative terms are not included. 
Generally, in MHD we introduce a Laplacian operator for both the velocity and magnetic field equations in order to mimic in a simple way the 
mechanism of dissipation which involves kinetic effects \citep{Belmont}. 
In MHD turbulence these terms are fundamental to assure the existence 
of a well resolved inertial range \Final{in which the dissipation (and forcing) is negligible}. 
In Hall MHD turbulence this modeling is \Final{less relevant} specially for the magnetic field because the nonlinear Hall term involves a 
derivative of order two \Final{and because the magnetic spectrum is generally steeper than in MHD. Then, with a classical Laplacian the dissipation is 
generally less localized at small-scales than in MHD which may reduce the size of the inertial range.} 
This means that to get an inertial range well separated from the \Final{small-scale} dissipative range \Final{it is better to have a dissipative term with} 
an order of derivative higher than two (\eg a bi-Laplacian). 

\Final{A detailed discussion of the dissipation mechanism is beyond the scope of this paper because, $(i)$ it depends on the application 
from which our derivation is independent, and $(ii)$ there are a variety of possible sources.}
\Final{However, it is relevant to make a comparison with 
the solar wind where a lot of studies are currently made. The weakly collisional plasma conditions in the solar wind imply that 
the mechanism responsible for the dissipation of turbulent fluctuations are likely to be of a kinetic nature like the damping of linear waves 
\cite{TenBarge2013}. This damping corresponds certainly to a term different from a Laplacian or double Laplacian. If the damping is 
either localized at some scales or significantly smaller in amplitude than the nonlinear fluctuations, then it will not affect significantly the inertial 
range and therefore our conclusion. Note, however, that the question of the relative importance of the damping is currently actively 
debated. For example, in a more recent paper \cite{Matthaeus2014} it is claimed that the solar wind data are compatible with the conclusion that the 
nonlinear time-scale is comparable or shorter than the typical time-scale of wave damping which renders the linear treatment questionable.}

\Final{The other important term for this type of analysis is the external forcing. In our case, we are mainly interested in the magnetic and generalized helicities. 
The injection of magnetic helicity is evoked in different astrophysical contexts like the Sun where the source can be the shearing motions \cite{Demoulin2002}. 
It is also discussed in the context of the dynamo problem where an inverse cascade of helicity can produce (or regenerate) a large-scale magnetic field 
\cite{pouquet76,Biskamp2003}. Magnetic turbulence in astrophysical systems is often governed by two effects, buoyancy and rotation, which naturally lead to helical 
flows and twisted field lines. The result is the production of a net magnetic helicity, and also probably some generalized helicity (since it contains the 
magnetic and kinetic helicities). This forcing is often made in a limited range of scales. For example, in the case of the geo-dynamo the typical scale ($\sim 10^3$\,km) corresponds 
roughly to the size (a fraction of the thickness) of the convection layer.}

\section{Exact Relation for Magnetic Helicity Conservation}\label{sec3}

When a flux of magnetic helicity is injected into a turbulent plasma, at a typical wave number $k_f$, an inverse cascade is expected. 
This happens only at MHD scale if $k_f d_i <1$ \citep{pouquet76} or at sub-ionic Hall MHD scales if $k_f d_i >1$ \citep{cho15,Galtier15} with eventually 
the possibility to extend the inverse cascade to MHD scales. 
In this section, we shall consider a forcing at intermediate scale such that $k_f d_i >1$ and focus the analysis on the inertial range at a scale $k \ll k_f$ (small-scale forcing). 
We see immediately a potential problem to derive a (statistically) stationary law in a finite size system: we need a dissipation at large-scale to counter balance the injection 
of magnetic helicity and avoid the formation of a condensate. 
In two-dimension hydrodynamic turbulence a similar situation is found but with an inverse cascade of kinetic energy \citep{Kraichnan67}. A frictional dissipation, 
usually due to \eg friction between the fluid and substrate, is then introduced into the system either numerically or/and analytically to avoid the formation of a condensate \citep{Chertkov07}. 
We will follow this line and assume the presence of some large-scale dissipation whose origin may depend on the problem (\eg the wall in a dynamo experiment). 

\Final{Note that in the case of the solar wind, since Hall MHD contains the MHD scales, an inverse cascade of magnetic helicity can continue 
potentially up to the largest scales of the system, \ie to frequencies much lower than the cyclotron frequency. However, another limitation 
exists: the magnetic helicity is not conserved in MHD if a uniform magnetic field is present \cite[see \eg][]{Shebalin2006}, and the solar wind is 
composed of a large-scale magnetic field plus fluctuations. In other words, when the inverse cascade crosses the scale $\sim d_i$ the transfer 
gets worse and we have therefore a nonlinear (non-dissipative) limit in length scale to the inverse cascade.}

Unlike \citep{PGP2003} we define the {\it symmetric} two-point correlation function associated with the magnetic helicity as
\begin{equation}
R_H = R'_H = \left\langle  \frac{\aa \cdot \bb'  + \aa' \cdot \bb}{2} \right\rangle , 
\end{equation}
where the primed and unprimed quantities correspond to the points $\xx'$ and $\xx$ respectively, with $ \xx' = \xx + \rr$, and $\langle \rangle$ means an 
ensemble average (which is equivalent by ergodicity to a spatial average in homogeneous turbulence). 
Using Eqs. \eqref{first2} and \eqref{second2}, and defining $\UU = \vv - d_i \jj $, we can write 
\begin{align}
&\p_t \left( {R_H + R'_H} \right)  \label{sixth} \\
&= \left\langle \aa' \cdot \p_t \bb + \bb \cdot \p_t \aa' + \aa \cdot \p_t \bb' + \bb' \cdot \p_t \aa \right\rangle  \nonumber \\
&= \left\langle \aa' \cdot \left[ \nabla \times \left( \UU \times \bb \right) \right] 
+ \bb \cdot \left( \UU' \times \bb' \right) + 2 \bb \cdot  \nabla' \psi' \right\rangle + {\cal D_H} \nonumber \\
&+ \left\langle \aa \cdot \left[ \nabla' \times \left( \UU' \times \bb' \right) \right] 
+ \bb' \cdot \left( \UU \times \bb \right)  + 2 \bb' \cdot  \nabla \psi  \right\rangle + {\cal F_H}   \nonumber \\ 
&= \left\langle \nabla \cdot \left[ \left( \UU \times \bb \right) \times \aa' \right] + \bb \cdot \left[ \UU' \times \bb' + 2  \nabla' \psi' \right]  \right\rangle + {\cal D_H} \nonumber \\
&+ \left\langle \nabla' \cdot \left[ \left( \UU' \times \bb' \right) \times \aa \right] + \bb' \cdot \left[ \UU \times \bb + 2  \nabla \psi\right]  \right\rangle + {\cal F_H}  ,  
\nonumber
\end{align}
with ${\cal D_H}$ a large-scale dissipation and 
\be
{\cal F_H} = \left\langle \aa' \cdot {\bf f_b} + \bb \cdot {\bf f'_a} + \aa \cdot {\bf f'_b} + \bb' \cdot {\bf f_a} \right\rangle ,
\ee
a small-scale forcing. By using the relation $ \bb = \nabla \times \aa $ and the statistical homogeneity, we obtain
\begin{align}
&\left\langle \nabla \cdot \left[ \left( \UU \times \bb \right) \times \aa' \right] +\nabla' \cdot \left[ \left( \UU' \times \bb' \right) \times \aa \right] \right\rangle 
\nonumber \\ 
&= - \left\langle \nabla' \cdot \left[ \left( \UU \times \bb \right) \times \aa' \right] + \nabla \cdot \left[ \left( \UU' \times \bb' \right) \times \aa \right] 
\right\rangle \nonumber \\
&= \left\langle \left( \UU \times \bb \right) \cdot \bb' + \left( \UU' \times \bb' \right) \cdot \bb \right\rangle . \label{seventh}
\end{align}
Inserting this above simplification in Eq. \eqref{sixth}, we find
\begin{align}
& \p_t   \left( \frac{R_H + R'_H}{2} \right) \nonumber \\
&= \left\langle \left[  \left( \UU \times \bb \right) + \nabla \psi\right] \cdot \bb' 
+ \left[ \left( \UU' \times \bb' \right) + \nabla' \psi' \right]  \cdot \bb \right\rangle \nonumber \\
&+ {{\cal D_H} \over 2} + {{\cal F_H} \over 2}  \nonumber \\
&= - \left\langle \delta  \left( \UU \times \bb \right)  \cdot \delta \bb \right\rangle +  {{\cal D_H} \over 2} + {{\cal F_H} \over 2} , \label{eighth}
\end{align}
where we use 
$$ 
\left\langle \nabla \psi \cdot \bb' + \nabla' \psi' \cdot \bb \right\rangle = - \left\langle  \psi (\nabla' \cdot \bb') +  \psi' (\nabla \cdot \bb) \right\rangle = 0
$$
due to statistical homogeneity. 
In the final step, we consider a stationary state (in the limit of infinite kinetic and magnetic Reynolds numbers) corresponding to the magnetic helicity conservation for which the left hand side of Eq. ({\ref{eighth}}) vanishes. 
Under this condition, we derive an expression for the inertial range within which the small-scale forcing term becomes negligible and the large-scale dissipative effect 
gives the mean magnetic helicity flux dissipation rate $\eta_M$ which is also equal to the mean magnetic helicity flux transfer rate. 
The final form of the exact relation is then given by 
\begin{equation}
\eta_M = \left\langle \delta \left( \UU \times \bb \right) \cdot \delta \bb \right\rangle ,  \label{premagnetic} 
\end{equation}
which can also be written as
\be
\eta_M  =  \left\langle \delta \left( \vv \times \bb \right)  \cdot \delta \bb \right \rangle 
- d_i \left\langle \delta \left( \jj \times \bb \right)  \cdot \delta \bb \right\rangle . \label{mag2}
\ee
In the large-scale MHD limit ($k d_i \ll 1$) we may recover the result obtained by \cite{PGP2003} when it is symmetrized. 
In the small-scale electron MHD limit ($k d_i \gg 1 $ or equivalently $\vv \to 0$) the above relationship is further simplified to 
\begin{equation}
\eta_M = d_i \left\langle \delta ( \bb \times \jj )  \cdot \delta \bb \right\rangle . \label{limit}
\end{equation}
Equation (\ref{mag2}) is the first main result of this paper. 
It is an exact relation valid for three-dimensional homogeneous Hall MHD turbulence without the assumption of isotropy. 
\Final{It is relevant to make a comparison with previous results and in particular with the magnetic helicity law derived in MHD \citep{PGP2003} which is the 
large-scale limit of Hall MHD. Unlike \citep{PGP2003} the assumption of isotropy is not made in our derivation giving to the new law a broader application like \eg to 
space plasmas where turbulence is rarely isotropic \cite[see \eg][]{Stawarz09}. Additionally, the law (\ref{mag2}) implies products of two increments (\ie only fluctuations), 
a situation well adapted to turbulence whereas the final law in \citep{PGP2003} did not consist of two-point increments.} 
\Final{A comparison with the classical derivation of the four-thirds laws for the 
energy \citep[see \eg][]{PP1998b} reveals another difference: our derivation does not lead to the appearance of a divergence operator which renders difficult the evaluation 
of the energy cascade rate in general (\ie when isotropy is not assumed). In our case, we see that the evaluation of $\eta_M$ requires only the computation of a 
scalar product of two increments. This property should be helpful for the study of the magnetic helicity effects in space (anisotropic) plasmas like \eg in the solar wind 
\cite{howes10}.}
Finally, from expressions (\ref{mag2}) and (\ref{limit}), one can easily infer that both in Hall MHD and electron MHD no turbulent flux of magnetic helicity exists if the 
system satisfies the corresponding Beltrami conditions \ie $\UU \parallel \bb$ and $ \jj \parallel \bb $ respectively. 
\Final{Therefore, we recover well} the theory of plasma relaxation to Beltrami alignment \citep{Yoshida02}. 

A power law spectrum in $-2$ is expected for the magnetic helicity when an inverse cascade happens at MHD scales \cite{pouquet76} (for simplicity, 
isotropy is assumed for the discussion). Dimensionally (assuming simply the maximal helicity state), this scaling corresponds to a $-1$ power law for the energy and thus 
to a constant $\delta b$. The exact relation (\ref{mag2}) tells us at the dimensional level (for simplicity we do not consider the effect of the scalar product which may give an 
additional scaling dependence) that necessarily the other increment $\delta (\vv \times \bb)$ has no scale dependence to keep globally constant the right-hand side term. 
The situation is different at sub-ion electron MHD scales where the inverse cascade of magnetic helicity may lead to (by using as before a Kolmogorov phenomenology 
with the assumption of maximal helicity state but in which the transfer time is now $\sim r^2/(d_i b)$) a corresponding power law spectrum in $-8/3$. Hence, a $-5/3$ magnetic 
energy spectrum and therefore $\delta b \sim r^{1/3}$. The exact relation (\ref{limit}) tells us that in this case necessarily we have dimensionally 
$\delta (\bb \times \jj) \sim r^{-1/3}$. A possible interpretation of this scaling law is that a phenomenon of alignment between $\bb$ and $\jj$ appears when one goes from 
small- to large electron MHD scales. (Note that in two dimensions this conclusion may change \cite{shaikh05}.)

\section{Exact Relation for Generalized Helicity Conservation}
In this section, we shall derive a universal law for the generalized helicity $H_G$. Hereinafter, we shall denote ${\bf \Upsilon} = \left( \aa + d_i \vv \right)$ 
the generalized vector potential whereas ${\bf \Omega}$ is the generalized vorticity. 
The generalized vorticity ${\bf \Omega}$ is somewhat analogous to the magnetic field in standard MHD \citep{woltjer58}. 
Indeed, both quantities obey the same Lagrangian equation which is for ${\bf \Omega}$ 
\be
\frac{d {\bf \Omega}}{dt} = {\bf \Omega} \cdot \nabla \, \vv \, . 
\ee
By applying the Helmholtz's law to Hall MHD, we see that the generalized vorticity lines are frozen into the plasma. 
However, the fact that at small-scales $\vv \times \bb - d_i \jj \times \bb \simeq {\bf v_e} \times \bb$, leads to 
\be
\frac{d \bb}{dt} = \bb \cdot \nabla \, {\bf v_e} \, , 
\ee
which shows that the magnetic field, which is no longer frozen in the fluid due to the Hall term, is still frozen with respect to the electronic fluid. 
Like the magnetic helicity in incompressible MHD, it is possible to show that the generalized helicity of a generalized vorticity tube is conserved over time \citep{turner86}. 
Therefore, $H_G$ provides a measure of the degree of structural complexity (the topology) of an incompressible Hall MHD flow.

With our notation, the generalized helicity can be written as $H_G =  {\bf \Upsilon} \cdot {\bf \Omega}$. Using the above Eqs. \eqref{first1}--\eqref{second2}, we obtain
\begin{align}
\p_t {\bf \Upsilon}  &= \vv \times {\bf \Omega} - \nabla P_G + {\bf f}_{\Upsilon} , \\
\p_t {\bf \Omega} &= \nabla \times ( \vv \times {\bf \Omega} ) + {\bf f}_{\Omega} ,
\end{align}
where $P_G=d_i v^2/2 + d_i P -2 \psi$ is a generalized pressure and ${\bf f}_{\Upsilon,\Omega}$ denote the corresponding forcing terms. 
(The dissipative terms will be introduced later.) Now we construct the symmetric two-point correlation function for the generalized helicity, namely
\begin{equation}
R_G = R'_G= \left\langle  \frac{{\bf \Omega} \cdot {\bf \Upsilon}'  + {\bf \Omega}' \cdot {\bf \Upsilon}}{2} \right\rangle \, . 
\end{equation}
Below, we shall derive the evolution equation of the correlation function. For the same reasons as before we do not introduce dissipative terms. 
Only external forcing terms (${\cal F}_G$) are introduced in the equations; then we get
{\begin{widetext}
\begin{align}
\p_t (R_G + R'_G) 
&= \left\langle \left[ - \nabla P_c + \vv \times {\bf \Omega} \right] \cdot {\bf \Omega}' + {\bf \Upsilon} \cdot \left[ \nabla' \times \left( \vv' \times {\bf \Omega}' \right) \right]  \right\rangle  
+ \left\langle \left[ - \nabla' P'_c + \vv' \times {\bf \Omega}' \right] \cdot {\bf \Omega} + {\bf \Upsilon}' \cdot \left[ \nabla \times \left( \vv \times {\bf \Omega} \right) \right]  \right\rangle + {\cal F}_G  \nonumber \\
&= \left\langle \left( \vv \times {\bf \Omega} \right) \cdot {\bf \Omega}' + \left( \vv' \times {\bf \Omega}'  \right) \cdot {\bf \Omega} \right\rangle  
+ \left\langle \nabla' \cdot \left[ \left( \vv' \times {\bf \Omega}' \right) \times {\bf \Upsilon} \right] + \nabla \cdot \left[ \left( \vv \times {\bf \Omega} \right) \times {\bf \Upsilon}' \right] \right\rangle + {\cal F}_G , \label{sixthM}
\end{align}
\end{widetext}}

\noindent
where ${\cal F}_G = \langle {\bf \Omega} \cdot {\bf f}'_\Upsilon + {\bf \Upsilon}' \cdot {\bf f}_\Omega + 
{\bf \Omega}' \cdot {\bf f}_\Upsilon + {\bf \Upsilon} \cdot {\bf f}'_\Omega \rangle$.
The definition $ {\bf \Omega} = \nabla \times {\bf \Upsilon} $ and the assumption of statistical homogeneity lead to the relation
\begin{align}
&\left\langle \nabla' \cdot \left[ \left( \vv' \times {\bf \Omega}' \right) \times {\bf \Upsilon} \right] + \nabla \cdot \left[ \left( \vv \times {\bf \Omega} \right) \times {\bf \Upsilon}' \right] \right\rangle \nonumber \\ 
&= - \left\langle \nabla \cdot \left[ \left( \vv' \times {\bf \Omega}' \right) \times {\bf \Upsilon} \right] + \nabla' \cdot \left[ \left( \vv \times {\bf \Omega} \right) \times {\bf \Upsilon}' \right] \right\rangle \nonumber \\
&= \left\langle \left( \vv' \times {\bf \Omega}' \right) \cdot {\bf \Omega} + \left( \vv \times {\bf \Omega} \right) \cdot {\bf \Omega}' \right\rangle \, . 
\end{align}
Substituting this expression into Eq. \eqref{sixthM}, we get 
\begin{align}
\p_t  \left( \frac{R_G + R'_G}{2} \right) &= 
\left\langle \left( \vv' \times {\bf \Omega}' \right) \cdot {\bf \Omega} + \left( \vv \times {\bf \Omega} \right) \cdot {\bf \Omega}' \right\rangle + {{\cal F}_G \over 2} 
\nonumber \\
&= - \left\langle \delta \left( \vv \times {\bf \Omega} \right)  \cdot \delta {\bf \Omega} \right\rangle + {{\cal F}_G \over 2} \, ,
\end{align}
as $ \left( \vv \times {\bf \Omega} \right) \cdot {\bf \Omega} = \left( \vv' \times {\bf \Omega}' \right) \cdot {\bf \Omega}' = 0 $. 

For the final step of the development we shall introduce a dissipative term which is necessary to make the assumption of stationary state corresponding to 
the conservation of the generalized helicity. By construction, the generalized vorticity and the vector potential are made of two type of terms: a magnetic term 
and a kinetic term. That means we can distinguish two situations: the case where the magnetic energy ${\cal E}^m$ is much greater than the kinetic energy
${\cal E}^u$, and the opposite situation \ie ${\cal E}^u \gg {\cal E}^m$. In the first case the generalized helicity identifies to the magnetic helicity and as it 
was explained in the previous section we may expect an inverse cascade. This situation corresponds actually (at sub-ion scales) to electron MHD. 
In the second case, the generalized helicity identifies with the kinetic helicity. In this case the Hall MHD equations can be simplified to the Navier-Stokes 
equations for which the kinetic helicity is indeed an invariant that leads to a direct cascade \cite{chen1}. Therefore, we see that both a direct and an inverse 
cascade may happen for the generalized helicity. Note that this conclusion may also be reached when one uses a Gibbs ensemble analysis \citep{servidio08}. 
To establish an exact relation valid well inside the inertial zone we shall introduce a large or small-scale dissipation which allows us to assume a stationary 
state. In practice, that corresponds to a small or large-scale forcing respectively. 
We introduce the mean generalized helicity flux dissipative rate $\eta_G$ and obtain after simplification
\begin{equation}
\pm \eta_G = \left\langle \delta \left( \vv \times {\bf \Omega} \right)  \cdot \delta {\bf \Omega} \right\rangle  \, , \label{generalized}
\end{equation}
where the sign $+$ and $-$ correspond to a direct and inverse cascades respectively (by definition $\eta_G$ is positive). 

Expression (\ref{generalized}) is the second main result of this paper. It is an exact relation for three-dimensional homogeneous Hall MHD turbulence without the 
assumption of isotropy. When the magnetic energy dominates, we recover expression (\ref{mag2}) both in the large-scale and small-scale limits (in the latter case 
$\vv - d_i \jj \to {\bf v_e}$).

\section{Chirality of the helicity cascades}

\Final{Interestingly each helicity cascade (magnetic and generalized) can be associated to a specific polarization. A simple way to recognize this fact is to write 
the laws (\ref{mag2}) and (\ref{generalized}) for the helicities in terms of the following generalized vortices and velocities \cite{yoshida,Meyrand12}
\begin{align}
\bOmega_L &= \bb + d_i \nabla \times \vv \, , \\
\bOmega_R &= \bb \, , \\
\vv_L &= \vv \, , \\
\vv_R &= \vv - d_i \nabla \times \bb \, ,
\end{align}
where $L$ and $R$ indices refer to different polarities. 
To understand why these variables correspond to the $L/R$ polarities, one possibility is to study the exclusive linear effect of each variable in the absence of the other \cite{Meyrand12}. 
The Hall MHD equations can be written as
\be 
\partial_t \bOmega_{L/R} = \nabla \times (\vv_{L/R} \times \bOmega_{L/R}) \, . \label{toto}
\ee
When the equations are linearized, by imposing a uniform magnetic field $\bb_0$, we get
\be
\partial_t \tilde{\bOmega}_{L/R} = \nabla \times (\tilde{\vv}_{L/R} \times \bb_0) \, 
\label{LRequa}
\ee
where $\tilde{\bf a}$ denotes the linear (first order) perturbation of the variable {\bf a}. For the case where only the L-variables survive, we have ${\tilde{\bOmega}}_R = 0 = {\tilde{\vv}}_R$. Under this condition expression (\ref{LRequa}) reduces to
\begin{equation}
 d_i \partial_t (\nabla \times \tilde{\vv}) = \nabla \times (\tilde{\vv} \times \bb_0) \, . 
\end{equation}
For every perturbation, we assume plane wave solution of the form $ \exp{[i({\bf k} \cdot {\bf x} - \omega t)] }$ with angular frequency $\omega$ and wave vector ${\bf k}$. The above relation therefore transforms (with $ \partial_t \mapsto - i \omega$ and $ \nabla \mapsto i {\bf k}$) into
\begin{equation}
 \omega d_i ({\bf k} \times {\hat \vv}) = i b_0 k_{\parallel} {\hat \vv} , 
\end{equation}
where ${\hat {\bf a}}$ denotes the Fourier transform of the variable ${\bf a}$. This equation represents a left-circularly polarized wave and also necessarily gives $\omega = b_0 k_\parallel / (k d_i)$  which identifies the mode as known ion-cyclotronic mode. 
For the other case, we have similarly ${\tilde{\bOmega}}_L = 0 = {\tilde{\vv}}_L$ and the linearization of Eq. (\ref{toto}) gives
\begin{equation}
\partial_t {\tilde{\bb}} = - d_i \nabla \times [(\nabla \times {\tilde{\bb}}) \times \bb_0] \, .
\end{equation}
After Fourier transformation, this simply writes
\begin{equation}
 i \omega {\hat \bb} = - d_i b_0 k_{\parallel} ({{\bf k} \times \hat \bb}) \, ,
\end{equation}
which is the equation of a right circularly polarized wave mode. In addition, we also get $\omega =  b_0 k_\parallel k d_i$ thereby recognizing the well-known right-circularly polarized whistler mode.}

\Final{It is straightforward to see (by a simple substitution) that the exact relations (\ref{mag2}) and (\ref{generalized}) 
can be written only in terms of the generalized fields.
We find 
\begin{align}
\eta_M  &=  \left\langle \delta \left( \vv_R \times \bOmega_R \right)  \cdot \delta \bOmega_R \right \rangle , \label{HR} \\
\pm \eta_G &= \left\langle \delta \left( \vv_L \times \bOmega_L \right)  \cdot \delta \bOmega_L \right\rangle  \, . \label{HL}
\end{align}
The form of expressions (\ref{HR})--(\ref{HL}) demonstrates that in one hand the magnetic helicity cascade is by nature a process
implying only the right-handed fluctuations whereas the generalized helicity cascade implies only the left-handed fluctuations. While the former
property may be expected since the magnetic helicity is an ideal invariant of electron MHD, the latter is less trivial. Following a recent analysis
\cite{Meyrand12} we may call ion MHD the regime where only right fluctuations persist.}

\section{Discussion} \label{sec4}

\begin{figure}
\includegraphics[scale=0.6,trim = 35mm 0mm 35mm -5mm]{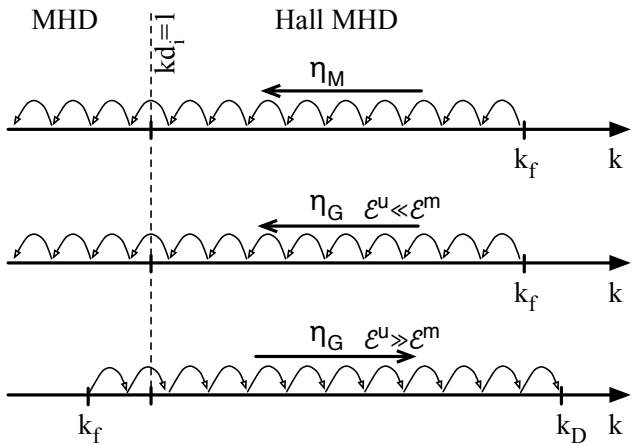}
\caption{Schematic view of the turbulent helicity cascades in Hall MHD. 
Top: inverse cascade of magnetic helicity with a small-scale forcing ($k_fd_i \gg 1$).
Middle: inverse cascade of generalized helicity in a magnetic regime (${\cal E}^m \gg {\cal E}^u$) and with a small-scale forcing ($k_fd_i \gg 1$).
Bottom: direct cascade of generalized helicity in a kinetic regime  (${\cal E}^u \gg {\cal E}^m$) and with a large-scale forcing ($k_fd_i < 1$).
The vertical dotted line ($k d_i =1$) separates the MHD scales (on the left) from the Hall MHD scales (on the right). $k_D^{-1}$ represents the dissipation length scale.}
\label{Fig1}
\end{figure}
\Final{Equations (\ref{mag2}), (\ref{generalized}) and (\ref{HR}), (\ref{HL}) are the main results of this paper. 
These exact relations are valid for homogeneous Hall MHD turbulence without the assumption of isotropy. Unlike the traditional Yaglom form, the non-isotropic expression does not involve any global divergence term and provides, therefore, a direct evaluation (using numerical simulations or observational data) of the transfer rates for the helicities. (Care should be taken for the fact that a stationary state corresponding to the helicity conservation is essential in order to verify these relations.)
In addition, for the first time we get the helicity laws for Hall MHD which are purely expressible in terms of two-point increments. We have also shown that the laws for the helicities can be written in terms of the generalized vortices and velocities which reveals the chiral properties of the helicity cascades -- the magnetic and generalized helicity cascades being associated to right and left handed fluctuations respectively. In addition, expressions (\ref{HR})--(\ref{HL}) show trivially that a Beltrami flow ($\bOmega_i = a_i \vv_i$), which is the state of maximum helicity, cannot produce a nonlinear cascade for the helicities. A summary of our results is proposed in Fig. \ref{Fig1}: in this fluid 
scenario, we do not introduce kinetic effects that should appear especially around the scale $k d_i \sim 1$. 
Note finally that expressions (\ref{mag2}), (\ref{generalized}) or (\ref{HR})--(\ref{HL}) can be used to evaluate experimentally the 
associated cascade rates. For example, the recently launched Magnetospheric Multiscale Mission (MMS) composed of four identically instrumented 
spacecrafts may provide this information during incursions into the solar wind by measuring the different increments at sub-ion scales (both for magnetic and plasma data).}

\bibliography{Banerjee_Galtier_Revision}
\end{document}